\documentclass[submission,copyright,creativecommons]{eptcs}
\providecommand{\event}{PxTP 2019} 
\usepackage{underscore}           

\usepackage{graphicx}
\usepackage{amsmath,amsthm,txfonts}
\usepackage{multirow}
\usepackage{bussproofs}
\usepackage{acronym}
\usepackage{forest}
\usepackage{footnote}
\usepackage{subfig}
\usepackage{enumitem}

\usetikzlibrary{arrows}

\usepackage[english, ruled, linesnumbered]{algorithm2e}
\SetKwProg{Fn}{Function}{}{}

\newtheorem{definition}{Definition}
\newtheorem{example}{Example}
\newtheorem{lemma}{Lemma}

\newtheorem{theo}{Theorem}

\title{Converting $\mathcal{ALC}$ Connection Proofs into $\mathcal{ALC}$ Sequents}
\author{Eunice Palmeira
\institute{Federal Institute of Alagoas\\ Macei\'{o} - AL, Brazil}
\email{eunicepalmeira@ifal.edu.br}
\and
Fred Freitas \qquad\qquad\qquad Jens Otten
\institute{ Federal University of Pernambuco \qquad University of Oslo\\Recife - PE, Brazil \quad\qquad\qquad\qquad Oslo, Norway}
\email{\qquad fred@cin.ufpe.br \qquad\qquad\quad jeotten@ifi.uio.no}
}
\def\titlerunning{Converting $\mathcal{ALC}$ Connection Proofs into $\mathcal{ALC}$ Sequents}
\def\authorrunning{E. Palmeira, F. Freitas \& J. Otten }

\begin{document}

\maketitle

\begin{abstract}
The connection method has earned good reputation in the area of automated theorem proving, due to its simplicity, efficiency and rational use of memory. This method has been applied recently in automatic provers that reason over ontologies written in the  description logic $\mathcal{ALC}$.  However,  proofs generated by connection calculi are difficult to understand. Proof readability is largely lost by the transformations to disjunctive normal form applied over the formulae to be proven. Such a proof model, albeit efficient, prevents inference systems based on it from effectively providing justifications and/or descriptions of the steps used in inferences. To address this problem, in this paper we propose a method for converting matricial proofs generated by the $\mathcal{ALC}$ connection method to $\mathcal{ALC}$ sequent proofs, which are much easier to understand, and whose translation to natural language is more straightforward. We also describe a calculus that accepts the input formula in a non-clausal $\mathcal{ALC}$ format, what simplifies the translation.
\end{abstract}

\section{Introduction}

Description Logics (DLs) \cite{Baader:1} are a family of knowledge representation formalisms considered as a fundamental foundation for the Semantic Web, as it constitutes the formalism underlying the Web Ontology Language (OWL) language. DL is an expressive, decidable subset of First Order Logic (FOL), successfully applied in several areas. DL provides a precise and unambiguous meaning to DL descriptions due to its formal semantics, and fast reasoners have been produced to the many fragments available \cite{Horrocks:1}. 

One of them, the $\mathcal{ALC}$ $\theta$-Connections Calculus, and its automated reasoner RACCOON (Reasoner based on the Connection Calculus Over ONtologies), is based on the Connection Method \cite{Freitas:1, Melo:1}, and was specifically developed to infer over the Description Logic $\mathcal{ALC}$ \cite{Freitas:1, Melo:1}. The calculus includes typical DL features and techniques, such as notation without variables, absence of Skolem functions/unification and, inclusion of a blocking rule to handle cycles, which guarantees termination to make for the case of cyclic ontologies. The Connection Calculus has earned good reputation in the area of automated theorem proving due to its simplicity, efficiency and rational use of memory. The method represents formulae as matrices, whose columns are conjunctive clauses; its proof procedure consists of horizontally traversing paths through the matrix in order to connect complimentary literals (e.g., $L$ with its complement $\neg L$). A pair $\lbrace L, \neg L \rbrace$, is called a connection, which corresponds to the validity the path being checked. Thus, a formula is valid if every path through the matrix corresponding to it has a connection.

Both calculi mentioned above, before attempting to find a proof, convert a formula into a disjunctive normal form. The translation to this clausal form often obscures the structure of the original formula and transforms some simple theorem proofs into difficult ones\cite{Plaisted:1}. In complex cases, the deductions' premise(s) and conclusion can no longer be clearly identified, once the transformation has been applied \cite{Bibel:2}. Thus, proof readability and understandability is largely lost, and consequently, it becomes quite difficult to provide justifications and/or descriptions of the steps used during inferences.

The $\theta$-Non-clausal $\mathcal{ALC}$ $\theta$-Connection Calculus is based on the $\mathcal{ALC}$ $\theta$-Connection Calculus and works directly on the structure of the original formula, thus avoiding the translation into a clausal form. Nevertheless, its proof format is still not intuitive, once, like other connection calculi, it consists of a set of complementary pairs found in each path through the matrix, when the formula is  valid.

The motivation of this work is to make a connection proof for $\mathcal{ALC}$ more readable so that, in a near future, justifications can be generated automatically in natural language. Therefore, this article proposes a conversion method that translates non-clausal $\mathcal{ALC}$ $\theta$-connection proofs into $\mathcal{ALC}$ sequent proofs. Sequent calculi have a more friendly proof representation than connection calculus; it conveys proofs in a formal logic argument style, where each proof line is a conditional tautology. Such translation should therefore contribute to a better user interaction with DL reasoners based on the Connection Method.

The DL $\mathcal{ALC}$ is presented in the next section; Section 3 brings an $\mathcal{ALC}$ non-clausal  Connection Calculus for $\mathcal{ALC}$; Section 4 introduces the $\mathcal{ALC}$ Sequent Calculus, to which proofs will be translated; the conversion process and its main concepts in Section 5; an overview of the main algorithms for the conversion method with its computational complexities in Section 6; and conclusions in Section 7.


\section{The Description Logic  \texorpdfstring{$\mathcal{ALC}$}{}}
\label{sec:TheDescriptionLogicALC}
An ontology \textit{O} in $\mathcal{ALC}$ is a set of axioms over a signature ($N_{C}, N_{R}, N_{O}$), where $N_{C}$ is \textit{the
set of concept names} (unary predicate symbols), $N_{R}$ is the \textit{set of role or property names} (binary predicate symbols); $N_{O}$ is the set of \textit{individual names} (constants) \cite{Baader:1}.
\textit{Concept expressions} are inductively defined as follows. $N_{C}$ includes $\top$, the \textit{universal
concept} that subsumes all concepts, and $\bot$, the \textit{bottom concept} subsumed by all concept names belong to $N_{C}$. If $r \in N_{R}$ is a \textit{role} and $C$, $D \in N_{C}$ are \textit{concepts}, then
th following formulae are also \textit{concepts}: (i) $C \, \sqcap \, D, \,$ (ii) $C \, \sqcup \, D, \,$ (iii) $\neg C, \,$ (iv)$\forall r.C$; (v) $\exists r.C$.

A knowledge base in DL consists of a set of basic axioms (TBox), and a set of axioms
specific to a particular situation (ABox). Two axiom types are allowed in a TBox $\mathcal{T}$: (i) $C \sqsubseteq D$; (ii) $C \equiv D$, standing for $C \sqsubseteq D$ and $D \sqsubseteq C$. An ABox $\mathcal{A}$ w.r.t. a TBox $\mathcal{T}$ is a finite
set of assertions of two types: (i) a \textit{concept assertion} is a statement of the form $C(a)$, where $a \in N_{O}$, $C \in N_{C}$ and (ii) a \textit{role assertion} $r(a,b)$, where $a, b \in N_{O}$, $r \in N_{R}$. An $\mathcal{ALC}$ formula is either an axiom or an assertion; an ontology \textit{O} is an ordered pair $(\mathcal{T},\mathcal{A})$. The semantics of concepts and ontologies is defined in the usual way - see, e.g., \cite{Baader:1}.

\section{The Non-clausal \texorpdfstring{$\mathcal{ALC}$}{} \texorpdfstring{$\theta$}{}-Connection Calculus}
\label{sec:TheNonClausalALCConnectionCalculus}



\begin{definition}\textbf{(Query).} A \textbf{query} $O \models \alpha$ is an $\mathcal{ALC}$ formula to be proven valid, where \textit{O} is an $\mathcal{ALC}$ ontology, and $\alpha$ is either a TBox or an ABox axiom to be proven a logical consequence from \textit{O}.
\end{definition}

\begin{definition} \textbf{(Literal, clause, matrix).} \textbf{$\mathcal{ALC}$ Literals} are atomic concepts or roles, possibly negated or instantiated in the form $L$ or $\neg L$. An $\mathcal{ALC}$ \textit{disjunction} is either a literal \textit{L}, a disjunction $(E_{0} \sqcup E_{1})$ or an universal restriction $\forall r.E_{0}$. An $\mathcal{ALC}$ \textit{conjunction} is either a literal \textit{L}, a conjunction $(E_{0} \sqcap E_{1})$ or an existential restriction $\exists r.E_{0}$, where $E_{0}$ and $E_{1}$ are expressions of arbitrary concepts (see DLs and its Mapping to FOL in  \cite{Baader:2}). \textbf{Clauses} are conjunctions of literals and matrices in the form $L_{1} \sqcap \ldots \sqcap L_{m}$, where each $L_{i}$ is a literal or a matrix.  A \textbf{matrix} of a formula (in DNF) is its representation as a set $\lbrace C_{1}, \ldots, C_{n}\rbrace $, where each $C_{i}$ is a clause.
\end{definition}

\begin{definition}\textbf{(Formula with polarity).} A \textbf{formula with polarity}, denoted by $F^{p}$, consists of a formula $F$ and a polarity $p$, where $p \in \lbrace 0, 1\rbrace$, that is, 0 is positive and 1 is negative. This concept is used to denote negation in a matrix, i.e. literals or matrices \textit{A} and $\neg A$ are represented by $A^{0}$ and $A^{1}$, respectively.
\end{definition}

\begin{definition}\textbf{($\mathcal{ALC}$ Non-Clausal Matrix).} An \textbf{$\mathcal{ALC}$ non-clausal matrix} is a set of clauses in which a clause is a set of literals and matrices. Let $F$ be a formula and \textit{p} be a polarity. The matrix of $F^{p}$, denoted by $M(F^{p})$, is inductively defined according to Table \ref{tab:NonClausalMatrix}, which indicates how the polarity is inherited by the (sub-)matrices of an $F^{p}$. The matrix of $F^{p}$ is the matrix $M(F^{0})$. Literals or (sub-)matrices involved in a universal restriction ($\forall r.C$) or in an existential restriction ($\exists r.C$) are underlined in the matrix.
\end{definition}

\begin{table}[ht]
\caption{Matrix of an $\mathcal{ALC}$ formula $F^{p}$.}
\label{tab:NonClausalMatrix}
\centering
\begin{tabular}{lllllll}

\cline{1-3}\cline{5-7}
Type & $F^{p}$ & $M(F^{p})$    &  & Type & $F^{p}$ & $M(F^{p})$ \\
\cline{1-3}\cline{5-7}   

Atomic & $A^{0}$ & $\lbrace\lbrace A^{0} \rbrace\rbrace$ & & $\beta$ & $(C \sqcap D)^{0}$ &  $\lbrace\lbrace M(C^{0}), M(D^{0})\rbrace\rbrace$\\

  & $A^{1}$ & $\lbrace\lbrace A^{1} \rbrace\rbrace$ & &  & $(C \sqcup D)^{1}$ &  $\lbrace\lbrace M(C^{1}), M(D^{1})\rbrace\rbrace$\\

$\alpha$  & $(\neg C)^{0}$ & $M(C^{1})$ & &  & $(C \sqsubseteq D)^{1}$ &  $\lbrace\lbrace M(C^{0}), M(D^{1})\rbrace\rbrace$\\

  & $(\neg C)^{1}$ & $M(C^{0})$ & & $\gamma$ & $(\forall rD)^{1}$ & $\lbrace\lbrace M(\underline{r^{0}}), M(\underline{D^{1}})\rbrace\rbrace$\\ 

  & $(C \sqcap D)^{1}$ &  $\lbrace\lbrace M(C^{1})\rbrace, \lbrace M(D^{1})\rbrace\rbrace$ & &  & $(\exists rD)^{0}$ &  $\lbrace\lbrace M(\underline{r^{0}}), M(\underline{D^{0}})\rbrace\rbrace$\\ 

  & $(C \sqcup D)^{0}$ &  $\lbrace\lbrace M(C^{0})\rbrace, \lbrace M(D^{0})\rbrace\rbrace$ & & $\delta$ & $(\forall rD)^{0}$ & $\lbrace\lbrace M(\underline{r^{1}})\rbrace, \lbrace M(\underline{D^{0}})\rbrace\rbrace$ \\ 
  
  & $(C \sqsubseteq D)^{0}$ &  $\lbrace\lbrace M(C^{1})\rbrace, \lbrace M(D^{0})\rbrace\rbrace$ & &  & $(\exists rD)^{1}$ &  $\lbrace\lbrace M(\underline{r^{1}})\rbrace, \lbrace M(\underline{D^{1}})\rbrace\rbrace$ \\
  
  & $(C \models D)^{0}$ &  $\lbrace\lbrace M(C^{1})\rbrace, \lbrace M(D^{0})\rbrace\rbrace$ & &  &  &     
\end{tabular}
\end{table}

\begin{definition} \textbf{(Positive) Graphical Representation of the Matrix).} In the \textbf{(positive) graphical representation of a matrix}, its clauses are arranged horizontally, while the literals and (sub-)matrices of each clause are arranged vertically. The restrictions are represented by solid lines; when a restriction involves more than one clause, its literals are indexed in the bottom with the same index in the matrix column in the written representation, for example, the notation $L_{i}$ (see example \ref{ex:QueryALC01}); restrictions with indexes are represented with horizontal lines; restrictions without indexes with vertical lines.
\end{definition}

\begin{example}\label{ex:QueryALC01}(Query, clause, $\mathcal{ALC}$ non-clausal matrix, formula with polarity, graphical representation of a matrix). The query $F_{1} = \lbrace\exists hasPet.Cat \sqsubseteq CatOwner,$ $OldLady \sqsubseteq \exists hasPet.Animal \sqcap \forall hasPet.Cat\rbrace\models OldLady \sqsubseteq CatOwner$ is read in FOL as:
\begin{equation}
\left.\begin{matrix}
\forall x ((\exists y \ hasPet(x,y)\wedge Cat(y)) \rightarrow  CatOwner(x)) & \\ 
\forall z (OldLady(z) \rightarrow \exists v (hasPet(z,v) \wedge Animal(v))) & \\
\wedge \forall k (hasPet(z,k) \rightarrow Cat(k))) \\
\end{matrix}\right\}
\models \forall u (OldLady(u) \rightarrow CatOwner(u)) \nonumber
\end{equation}
\end{example}
and is represented by the FOL matrix (\textit{a} is a Skolem terms, \textit{f} a function symbol):
{\small
\begin{eqnarray}\label{eq:CC-MatrizLPOALCFormula01}
\lbrace\lbrace hasPet(x,y), Cat(y), \neg CatOwner(x) \rbrace, \lbrace OldLady(z),\lbrace \lbrace \neg hasPet(z,f(z)) \rbrace,\lbrace 
\neg Animal(f(z))\rbrace, \lbrace  hasPet(w,k), \nonumber\\ 
\neg Cat(k) \rbrace\rbrace\rbrace,
\lbrace \neg OldLady(a)\rbrace, \lbrace CatOwner(a) \rbrace\rbrace \nonumber
\end{eqnarray}
}
and by the following $\mathcal{ALC}$ non-clausal matrix $M_{1}$, which is defined according to \ref{tab:NonClausalMatrix} (column indices relate the two clauses involved in a same restriction; variables are omitted as they are specified implicitly):
{\small
\begin{eqnarray}\label{eq:CCNCALC-SimplificadaNaoClausalFormula}
\lbrace \lbrace \underline{hasPet^{0}}, \underline{Cat^{0}}, CatOwner^{1}\rbrace, \lbrace OldLady^{0}, \lbrace\lbrace \underline{hasPet^{1}_{1}}\rbrace, \lbrace \underline{Animal^{1}_{1}}\rbrace, \lbrace\underline{hasPet^{0}}, \underline{Cat^{1}} \rbrace\rbrace \rbrace, \nonumber \\
\lbrace OldLady(a)^{1}\rbrace, \lbrace CatOwner(a)^{0}\rbrace \rbrace \nonumber
\end{eqnarray}
}
So, the graphical representation of $M_{1}$ is:
{\footnotesize
$$\begin{bmatrix} 
 		\begin{bmatrix}
			\left.\begin{matrix}
			hasPet^{0} \,\\ 
			Cat^{0}
			\end{matrix}\right|\\ 
			CatOwner^{1}
		\end{bmatrix} 
 		\begin{bmatrix} OldLady^{0} \\ 
 			\begin{bmatrix} [\underline{hasPet^{1}_{1}] [Animal^{1}_{1}}]      
 			\begin{bmatrix}
				\left.\begin{matrix}
					hasPet^{0} \, \\ 
					Cat^{1}
				\end{matrix}\right|
			\end{bmatrix} 
 		  \end{bmatrix}   
	    \end{bmatrix} 	   			
 		[ OldLady(a)^{1} ] [ CatOwner(a)^{0} ] 
\end{bmatrix}$$}

 Matrices of the form $M=\lbrace\ldots,\lbrace C_{1},\ldots, C_{n}\rbrace,\ldots\rbrace$ can be simplified to $M'=\lbrace\ldots, C_{1},\ldots,$ $C_{n},\ldots\rbrace$, where $C_{1},\ldots, C_{n}$ are clauses. 

 Clauses of the form $C=\lbrace\ldots,\lbrace$ $M_{1}$,$\ldots$, $M_{m}\rbrace$,$\ldots\rbrace$ can be simplified to $C'=\lbrace\ldots, M_{1},\ldots, M_{m},\ldots\rbrace$, where $M_{1},\ldots, M_{m}$ are matrices.

\begin{definition} \textbf{(Path).} 
A \textbf{path} through a matrix $M = \lbrace C_{1}, \ldots , C_{n} \rbrace$ is a set of literals containing a literal $L_{i}$ of each clause $ C_{i} \in M$, i.e., $\bigcup^{n}_{i=1} \lbrace L_{i} \rbrace$ with $L_{i} \in C_{i}$. A path through a matrix M (or a clause C) is inductively defined as follows. The (only) path through a literal L is $\lbrace L \rbrace$. If $p_{1},\ldots,p_{n}$ are paths through the clauses $C_{1},\ldots,C_{n}$, respectively, then $p_{1} \cup \ldots \cup p_{n}$ is a path through the matrix $M = \lbrace C_{1},\ldots,C_{n}\rbrace $. If $p_{1},\ldots,p_{n}$ are paths through the matrices/literals $M_{1},\ldots,M_{n}$, respectively, then $p_{1}, \ldots, p_{n}$ are also paths through the clause $C = \lbrace M_{1}, \ldots, M_{n}\rbrace$.
\end{definition}

\begin{definition} \textbf{(Connection, $\theta$-substitution, $\theta$-complementary connection).} A \textbf{connection} is a pair of literals $\lbrace E, \neg E\rbrace$ with the same concept/role name, but different polarities. A
\textbf{$\theta$-\textit{substitution}} assigns to each (possibly omitted) variable an individual or another variable (in the whole matrix). A \textbf{$\theta$-\textit{complementary connection}} is a pair of $\mathcal{ALC}$ literals $\lbrace E(x), \neg E(y)\rbrace$ or $\lbrace p(x, v), \neg p(y, u)\rbrace$, with $\theta (x) = \theta(y), \theta (v) = \theta (u)$. The complement $\overline{L}$ of a literal $L$ is $E$ if $L = \neg E$, and it is $\neg E$ if $L = E$.
\end{definition}

Simple term unification without Skolem functions is used to calculate $\theta$-substitutions. The application of a $\theta$-substitution to a literal is an application to its variables, i.e. $\theta(E) = E(\theta (x))$ and $\theta (r) = r(\theta (x), \theta(y))$, where \textit{E} is an atomic concept and \textit{r} is a role. Furthermore, $x^{\theta} = \theta(x)$.

\begin{example}\textbf{(Path, Connection, $\theta$-substitution, $\theta$-complementary connection).}
In the matrix $M_{1}$ of Example \ref{ex:QueryALC01}, $\lbrace hasPet^{0}\mid$, $\underline{hasPet_{1}^{1}}$, $\underline{Animal_{1}^{1}},$ $hasPet^{0}\mid$, $OldLady(a)^{1}$, $CatOwner(a)^{0} \rbrace$ and $\lbrace Cat^{0}$, $\underline{hasPet_{1}^{1}}$, $\underline{Animal_{1}^{1}}$, $Cat^{1}\mid$, $OldLady(a)^{1}$, $CatOwner(a)^{0} \rbrace$ are some paths through $M_{1}$. $\lbrace Cat^{0}\mid , Cat^{1}\rbrace$ is a connection. $\theta(OldLady^{0}) =$ $OldLady(\theta(y))^{0}$ and $\theta (hasPet^{0}) =$ $hasPet( \theta(y),x)^{0}$, where $\theta(y) = a$, are examples of $\theta$-substitution, and $\lbrace OldLady^{0}, OldLady(a)^{1}\rbrace$ is a $\theta$-complementary connection,
\label{ex:CC-SubstituicaoALC}
\end{example}

\begin{definition}\textbf{(Set of concepts, Skolem condition).} The \textbf{set of concepts} $\tau(x)$ of a variable or individual $x$ contains all concepts that were substituted/ instantiated by \textit{x} so far, i.e. $\tau(x) \stackrel{\text{def}}{=} \lbrace E(x) \in Path\rbrace$, where $E$ is a concept and $E(x)$ is a substituted/instantiated literal coming from this concept. The \textbf{Skolem condition} ensures that at most one concept is underlined in the graphical matrix. The condition is formally stated as, $\forall a \vert \lbrace \underline{E^{i}(a)} \in Path \rbrace \vert \leq 1$, with $a$ a variable/individual, and $i$ a column index.
\end{definition}

\begin{definition}\textbf{($\alpha$-Related Clause).} Let $C$ be a clause in a matrix $M$ and $L$ be a literal in $M$. $C$ is $\alpha$-related to $L$, iff $M$ contains (or is equal to) a matrix $\lbrace C_{1},\ldots,C_{n} \rbrace$ such that $C = C_{i}$ or $C_{i}$ contains $C$, and $C_{j}$ contains $L$ for some $1 \leq i, j \leq n$ with $i \neq j$. $C$ is
\textbf{$\alpha$-related clause} to a set of literals $\mathcal{L}$, iff $C$ is $\alpha$-related to all literals $L \in \mathcal{L}$.
\end{definition}

\begin{example}\textbf{($\alpha$-Related Clause)} In the matrix of Example \ref{ex:QueryALC01}, $\lbrace\underline{ Animal^{1}_{1}} \rbrace$ is $\alpha$-related to $\lbrace hasPet^{0},  Cat^{1}\rbrace$.
\end{example}

\begin{definition}\textbf{(Parent Clause).} Let $M$ be a matrix and $C$ be a clause in $M$. The clause $C' = \lbrace M_{1},\ldots,M_{n} \rbrace$ in $M$ is called the \textbf{parent clause} of $C$ iff $C \in M_{i}$ for some $1 \leq i \leq n$.
\end{definition}

\begin{example}\textbf{(Parent Clause).} In Example \ref{ex:QueryALC01},
$\lbrace OldLady^{0}, \lbrace \lbrace \underline{hasPet^{1}_{1}} \rbrace, \lbrace \underline{Animal^{1}_{1}} \rbrace, \lbrace hasPet^{0}, Cat^{1}\rbrace \rbrace \rbrace$ is parent clause of $\lbrace \underline{ hasPet^{1}_{1}} \rbrace$.
\end{example}

\begin{definition}\textbf{(Extension Clause).} Let $M$ be a matrix and $P$ a path (be a set of literals). Then the clause $C$ in $M$ is an \textbf{extension clause} of $M$ with respect to $P$, iff either $C$ contains a literal of $P$, or $C$ is $\alpha$-related to all literals of $P$ occurring in $M$ and if $C$ has a parent clause, it contains a literal of $P$.
\label{def:CC-ClausulaExtensao}
\end{definition}

In the extension rule of the $\mathcal{ALC}$ $\theta$-Connection Calculus (\ref{sec:TheFormalCalculus}) the new subgoal clause (set of literals that need to be connected) is $C_{2} \setminus \lbrace L_{2}\rbrace$. In the non-clausal connection calculus the extension clause $C_{2}$ might contain clauses that are $\alpha$-related to $L_{2}$ and do not need to be considered for the new subgoal clause. Hence, these clauses can be deleted from the subgoal clause.
The resulting clause is called the $\beta$-clause of $C_{2}$ with respect to $L_{2}$.

\begin{definition}\textbf{($\beta$-Clause).} Let $C = \lbrace M_{1},\ldots,M_{n} \rbrace$ be a clause and $L$ be a literal in $C$. The \textbf{$\beta$-Clause} of $C$ with respect to $L$, denoted by $\beta$-Clause$_{L}(C)$, is inductively defined:
$$\beta\text{-Clause}_{L}(C) := \left\{
\begin{array}{lc}
C \setminus\left \{ L \right \} & \text{if } L \in C,\\ 
 M_{1},\ldots ,M_{i-1},\lbrace C^{\beta}\rbrace,M_{i+1},\ldots,M_{n} & \text{otherwise},
\end{array}
\right.$$
where $C' \in M_{i}$ contains $L$ and $C^{\beta} := \beta$-Clause$_{L}(C')$.
\label{def:CC-BetaClausula}
\end{definition}

\sloppy
\begin{example}\textbf{(Extension Clause, $\beta$-Clause).} In Example
\ref{ex:QueryALC01}, 
$C = \lbrace OldLady^{0}, \lbrace \lbrace \underline{hasPet^{1}_{1}} \rbrace,$  $\lbrace \underline{Animal^{1}_{1}} \rbrace, \lbrace \underline{hasPet^{0}},$ $\underline{Cat^{1}}\rbrace \rbrace \rbrace$ 
is an extension clause with respect to $p = \lbrace CatOwner(a)^{0}, \underline{Cat}^{0} \rbrace$, 
while the clause $\lbrace OldLady^{0},$ $\lbrace \lbrace \underline{hasPet^{1}_{1}} \rbrace, \lbrace \underline{Animal^{1}_{1}} \rbrace, \lbrace hasPet^{0} \rbrace \rbrace \rbrace$ 
is a $\beta$-Clause of $C$ with respect to $L = \underline{Cat^{1}}$.
\end{example}

\subsection{The Formal Non-Clausal \texorpdfstring{$\mathcal{ALC}$}{} \texorpdfstring{$\theta$}{}-Connection Calculus}
\label{sec:TheFormalCalculus}
%

Suppose we wish to entail if $O \models\alpha$ is valid using a direct
method, like the Connection Method (CM). By the Deduction Theorem [3], we must then prove directly if $O\rightarrow\alpha$, or, in other words, if $\lnot O\vee{\alpha}$ is valid. This opposes to classical refutation methods, like tableaux and resolution, which builds a proof by testing whether $O\cup\left\{ \lnot\alpha\right\}\models\ \bot$. Hence, in the CM, the whole knowledge base KB should be negated. Given $O=\left\{\alpha_1,\alpha_2,\ \ldots,\alpha_n\right\},\ \alpha_i$ being literal conjunctions in the clausal connection method, all (negated KB) formulae are converted to the Disjunctive Normal Form (DNF). A query then is the matrix $\lnot O\vee{\alpha}$ (i.e., $\lnot\alpha_1\vee\lnot\alpha_2\vee\ \ldots\vee\lnot\alpha_n\vee\alpha$) to be proven valid. In the non-clausal calculus, instead of having clauses only with literals, they can also contain matrices, and no conversion is needed. If every path contains a ($\theta$-complementary) connection (representing a subformula $A \sqcup \lnot A$ in a disjunction, what makes this disjunction valid), then the matrix is valid.

\begin{definition}\textbf{(Non-Clausal $\mathcal{ALC}$ $\theta$-Connection Calculus)} Figure \ref{fig:FormalNonClausalALCThetaConnectionCalculus} shows the rules of the formal non-clausal $\mathcal{ALC}$ $\theta$-connection calculus. Rules are applied bottom-up. The words of the calculus are tuples $C, M, Path$, where $C$ is a clause, $M$ is a matrix corresponding to query $O \models \alpha$ and $Path$ is a set of literals. $C$ is called the subgoal clause. $C_{1}$, $C_{2}$ and $C_{3}$ are clauses. The index $\mu \in \mathbb{N}$ of a clause $C^\mu$ denotes that $C^\mu$ is the $\mu$-th copy of clause $C$, increased when $Copy$ is applied for that clause (the variable $x$ in $C^\mu$ is denoted $x_\mu$). When $Copy$ is used, it has to be followed by the application of $Extension$ or $Reduction$, to avoid non-determinism in the rules’ application. The \textit{Blocking Condition} is defined as follows: the new individual $x^{\theta}_\mu$ (if it is new, then $x^{\theta}_\mu \not\in N_{O}$, as in the condition) is only created if the set of concepts of the previously created individual $\tau (x^{\theta}_{\mu -1})$ is not a subset of the set of concepts of the penultimate copied individual, i.e., $\tau (x^{\theta}_{\mu-1}) \not \subseteq \tau (x^{\theta}_{\mu-2})$.
\end{definition}

\begin{figure}[!ht]
{\footnotesize  
\begin{eqnarray}
Axiom (A)& & \frac{}{\lbrace\rbrace,M,Path} \nonumber\\ 
Start (S)& & \frac{C_{1}, M, \lbrace\rbrace}{\varepsilon,M,\varepsilon} \text{ with } C_{1} \in \alpha \nonumber\\
Reduction (R)& & \frac{C, M, Path \cup \lbrace L_{2} \rbrace}{C \cup \lbrace L_{1} \rbrace,M,Path \cup \lbrace L_{2} \rbrace} \nonumber\\
& & \text{\rm \ with } \theta(L_{1}) = \theta (\overline{L_{2}}) \text{\rm \ and\ the\ Skolem\ condition\ holds}  \nonumber\\
& &  \nonumber\\
Extension (E) & & \frac{C_{3}, M, Path \cup \lbrace L_{1} \rbrace \qquad C, M, Path}{C \cup \lbrace L_{1}\rbrace, M, Path} {\rm \ with}\ C_{3}:=\beta{\rm -\textit{clause}}_{L_{2}}(C_{2} ), \nonumber\\
& & C_{2} {\rm \ is\ an\ extension\ clause\ of}\ M {\rm \ wrt.} \ Path \cup \lbrace L_{1} \rbrace, \nonumber\\
& & L_{2} \in C_{2},\, \theta(L_{1}) = \theta (\overline{L_{2}}) {\rm \ and\ the\ Skolem\ condition\ holds} \nonumber\\
& &  \nonumber\\
Decomposition (D) & & \frac{C \cup C_{1}, M, Path}{C \cup \lbrace M_{1}\rbrace, M, Path} {\rm \ with}\ C_{1} \in M_{1}  \nonumber\\
& &  \nonumber\\
Copy (C)& & \frac{C \cup \lbrace L_{1}\rbrace, M \cup \lbrace C^{\mu}_{2} \rbrace, Path}{C \cup \lbrace L_{1}\rbrace, M, Path} {\rm \ with\ } C^{\mu}_{2} {\rm \ is \ a \ copy \ of\ } C_{1},\nonumber\\
& & L_{2} \in C^{\mu}_{2}, \; \theta(L_{1})=\theta(\overline{L_{2}}) {\rm \ and\ the\ blocking\ condition\ holds} \nonumber
\end{eqnarray}
\caption{Non-clausal $\mathcal{ALC}$ $\theta$-Connection Calculus.}
\label{fig:FormalNonClausalALCThetaConnectionCalculus}}
\end{figure}

The calculus consists of six rules. The Axiom, Start, Reduction and Copy rules are the same as the ones from the $\mathcal{ALC}$ $\theta$-Connection Calculus. The Extension rule was modified to contain a $\beta$-Clause and the Decomposition rule \cite{Otten:1} splits subgoal clauses into their sub-clauses.

\begin{lemma}\textbf{(Matrix characterization)}. A matrix M is valid iff there exist an index $\mu$, a set of $\theta$-substitutions $\langle \theta_i \rangle$ and a set of connections S, s.t. every path through $M^\mu$, the matrix with copied clauses, contains a $\theta$-complementary connection ${L_1^\theta,L_2^\theta}$ in S, i.e. a connection with $\theta\left(L_1\right)=\theta\left(\overline{L_2}\right)$. The tuple $\langle \mu,\langle \theta_i \rangle, S \rangle$ is called a matrix proof.
\end{lemma}

\begin{figure}[!htb]
\centering
\includegraphics [scale=0.6] {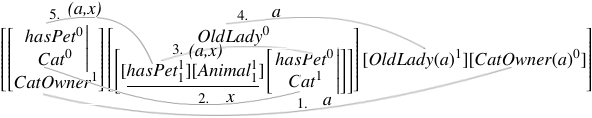}
\caption{The $\mathcal{ALC}$ non-clausal matrix proof of the $F_{1}$ using the graphical matrix representation.}
\label{fig:alcnonclausalconnectionproofcomplete}
\end{figure}

\begin{example}\textbf{(Non-Clausal $\mathcal{ALC}$ $\theta$-Connection Calculus)}. Figure \ref{fig:alcnonclausalconnectionproofcomplete} shows the proof for the $F_{1}$ of Example \ref{ex:QueryALC01} using the matrix representation. 

\end{example}

The proof starts (1) by choosing a clause from the consequent as the \textit{start clause}, in this case, $\lbrace CatOwner(a) \rbrace$, and a literal of that clause is selected, $CatOwner(a)^{0}$. This literal is connected to $CatOwner^{1}$ by an extension step and instance \textit{a} is the $\theta$-substitution of $CatOwner^{1}$ and $CatOwner(a)^{0}$. This connection is still not enough to prove all the paths starting from $CatOwner(a)^{0}$; the paths that start in it and pass through the literals from the other connected clause, namely, $Cat^{0}$ and $hasPet^{0}$, are still to be verified. Indeed, each connection creates two  sets of literals to be checked, the remaining literals from each of the clauses involved in the connection. 
In the new extension step (2), the connection $\lbrace Cat^{0}, Cat^{1}\rbrace$ is established on the variable (or fictitious individual) $x$, as it is not necessary yet to commit the substitution with an already existing individual. There is still remaining literals to be verified, the ones resulting from the clause to which $Cat^{0}$ belongs. Next (3), the $hasPet^{0}$ predicate is connected, and the $\theta$-substitution generates the pair \textit{(y,x)} (not shown in figure), for the connection. $OldLady^{0}$ is connected to $OldLady(a)^{1}$ (4), and then (5), when the connection $\lbrace hasPet^{0}, hasPet^{1} \rbrace$ is settled (using a reduction step, as there was already a connection with the same literal in the path), \textit{y}  was $\theta$-substituted by \textit{y} (i.e., $\theta(y) = a$), thus forming the pair \textit{(a,x)}. This $\theta$-substitution over \textit{y} is then propagated through the path. Since every path through $M_{1}$ contains a $\theta$-complementary connection, $F_{1}$ is valid. However, the readability of the proof is largely lost by the transformations applied on the formulas to be proven, making it difficult to translate the steps into natural language.

Next, we present the Sequent Calculus to which $\mathcal{ALC}$ non-clausal proofs will be translated.



\section{An \texorpdfstring{$\mathcal{ALC}$}{} Sequent Calculus}
\label{sec:SequentCalculusALCSubsumption}

According to \cite{Borgida:1}, sequent calculi axiomatizes the relation of logical consequence (entailment), and this has an obvious parallel with the relation of subsumption, which is a keystone for DL representation and calculi. Bearing this in mind, Borgida et al proposed a sequent calculus for subsumption inferences in $\mathcal{ALC}$ as an extension of the standard sequent calculus, in which there are no rules of implication, as they are indeed subsumption rules, so implication is replaced by $\vdash$ without loss of meaning. In their calculus, terms are not moved from one side to the other of the turnstile during the proof, thus preserving the structure of the original subsumption, and in the case of multiple subsumptions, parentheses help in identifying the main subsumptions. Because of that, additional rules were created in which the negation is inserted in front of each construct, thus eliminating negation rules  (l$\neg$, r$\neg$), what requires changing sequent antecedents to successors and vice versa. The calculus is divided in three parts: the first two describe sets of rules, while the last describes a set of axioms (see Figure \ref{fig:ALCSequentCalculusRules}, where $a$ and $b$ are arbitrary formulas and $X$ and $Y$ are arbitrary sequences of formulae).

\begin{itemize}

\item \textbf{Rules for propositional formulae:} rules $\sqcap$ and $\sqcup$ are duplicated by adding the negation rules for these connectives ($\neg \sqcap, \neg \sqcup$), while the proper negation rules ($\neg$) were modified to include the double negation rule ($ \neg \neg $);

\item \textbf{Rules for quantified formulae:} in \cite{Borgida:1}, modal formulae are used ($r\Box$, $l\Diamond$) and their negated rules ($l\neg\Box$, $r\neg\Diamond$). Here, we replace these rules by their equivalents ($r\forall$, $l\exists$) and ($l\neg\forall$, $r\neg\exists$).  The $\exists$\textit{-rules} are the dual $\forall$\textit{-rules}. A condition is explicitly considered for the application of these rules: the rule applies only if all homologous universal and existential formulae (e.g. $\forall h.C$ and $\exists h.C$ are homologous, $\forall h.C$ and $\exists f.C$ not) are joined together on the left and right sides of the sequent in the precondition. The rule is then applied only once;

\item \textbf{Termination axioms:} unlike the standard sequent calculus, there are six termination axioms; all of them can be reduced to $X, a \vdash a, Y$ by applying the rules. The application of the $\neg$\textit{-rules} forces formulae from the antecedent to the successor or vice versa, to be transformed until it gets to $X, a \vdash a, Y$, a procedure that is avoided in this calculus. Therefore, the additional termination axioms are necessary to ensure that formulae are never shifted from one side of the sequent to the other.
\end{itemize}

Although not stated explicitly, the calculus contains a cut rule, and the cut elimination theorem is valid in this case; it is stated below.

\begin{theo}\textbf{Cut Elimination Theorem \cite{Girard:1}}. Let $S$ be a set of sequents (axioms) and $s$ an individual sequent. $S \vdash_{SC} s$, if and only if, there is a proof in $SC$ of $s$ whose leaves are either logical or sequent axioms obtained by the substitution of $S$-belonging sequents, where the cut rule, $\frac{\Gamma \; \vdash \; \Delta, \, A \qquad A, \, \Sigma \; \vdash \; \Pi}{\Gamma, \, \Sigma \; \vdash \; \Delta , \, \Pi}$, is only applied with a premise being an axiom.

\label{prop:CutEliminationTheorem}
\end{theo}

\begin{figure}[!htb]
\centering
{\scriptsize
\begin{tabular}{clcl}
\multicolumn{2}{l}{{\small \textbf{Rules for propositional formulae}}} & & \\
{\large$\frac{X \, ,  a \, ,  b  \; \vdash \;  Y}{X \, , \, \, a \sqcap b  \; \vdash \;  Y}$} & $(l \sqcap)$ & {\large$\frac{X \; \vdash \; a \, , \, Y  \quad  X \; \vdash \; b \, , \, Y}{X \, , \; \vdash \; a \sqcap b \, , \, Y}$} & $(r \sqcap)$  \\
&&&\\
{\large$\frac{X \, , \neg a \; \vdash \; Y \quad X \, , \neg b \; \vdash \; Y}{X \, , \neg (a \sqcap b) \; \vdash \; Y}$} & $(l \neg \sqcap)$&{\large$\frac{X \; \vdash \; \neg a \, , \neg b \, , \; Y}{X \; \vdash \; \neg (a \sqcap b) \, , \; Y}$} & $(r \neg \sqcap)$ \\
&&&\\
{\large$\frac{X \, , \, \, a \; \vdash \; Y \quad X \, , \, \, b \; \vdash \; Y}{X \, , \, \, a \sqcup b \; \vdash \; Y}$} & $(l \sqcup)$ &  {\large$\frac{X \;  \vdash \; a\, , \, b \, , Y}{X \; \vdash \; a \sqcup b \, ,\; Y}$} & $(r \sqcup)$ \\
&&&\\
{\large$\frac{X \, , \neg a \, , \neg b \; \vdash \; Y}{X \, , \neg (a \sqcup b) \; \vdash \; Y}$} & $(l \neg \sqcup)$ & {\large$\frac{X \; \vdash \; \neg a \, , \, Y \quad X \; \vdash \; \neg b \, , \, Y}{X \,  \; \vdash \; \neg (a \sqcup b) \, , \, Y}$} & $(r \neg \sqcup)$ \\
&&&\\
{\large$\frac{X \, , \, \, a \; \vdash \; Y}{X \, , \neg \neg a \; \vdash \; Y}$} & $(l \neg \neg)$ & {\large$\frac{X \; \vdash \; a \, , \, Y}{X \; \vdash \; \neg \neg a \, , \, Y}$} & $(r \neg \neg)$ \\
&&&\\
\multicolumn{2}{l}{{\small \textbf{Rules for quantified formulae}}} & &  \\
{\large$\frac{X' \; \vdash \; b \, , \, Y'}{X \; \vdash \; \forall r.b \, , \;   Y}$} & $(r \forall)$ & {\large$\frac{X' \, , \, b \; \vdash \; Y'}{X \, , \; \exists r.b \; \vdash \; Y}$} & $(l \exists)$  \\
&&&\\
{\large$\frac{X'\, , \; \neg b \; \vdash \; Y'}{X \, , \, \neg \forall r.b \; \vdash \;  Y}$} & $(l \neg \forall)$ & {\large$\frac{X' \; \vdash \; \neg b \, , \, Y'}{X \; \vdash \; \neg \exists r.b \, , \, Y}$} & $(r \neg \exists)$  \\
&&&\\
\multicolumn{4}{c}{where $X' = \lbrace a \; | \; \forall r.a \; \in \; X\rbrace \cup  \lbrace \neg a \; | \; \neg \exists r.a \; \in \; X\rbrace $, and} \\
\multicolumn{4}{c}{\quad $Y' = \lbrace a \; | \; \exists r.a \; \in \; Y\rbrace \cup  \lbrace \neg a \; | \; \neg \forall r.a \; \in \; Y\rbrace $} \\
\multicolumn{2}{l}{{\small \textbf{Termination axioms}}} & &   \\
$X, \; a \; \vdash \; a \, , \;  Y$ &(=)&  $X \; , \neg a \; \vdash \; \neg a \, , \; Y$ &(=)\\ 
$X,\; a \, , \; \neg a \; \vdash \;  Y$ &(l$\uparrow$)&  $X \; \vdash \; a \, , \; \neg a \, , \;  Y$ &(r$\uparrow$)\\
$X \; , \perp \; \vdash \;  Y$ &(l$\perp$)& $X \; \vdash \; \top \,, \; Y$ &(l$\top$) \\
&&&\\
\multicolumn{2}{l}{{\small \textbf{Cut rule}}} & &   \\
\multicolumn{4}{c}{\large$\frac{\Gamma \; \vdash \; \Delta, \, A \qquad A, \, \Sigma \; \vdash \; \Pi}{\Gamma, \, \Sigma \; \vdash \; \Delta , \, \Pi}$}
\end{tabular}
}
\caption{The Sequent Calculus for $\mathcal{ALC}$ Subsumption \cite{Borgida:1}.}
\label{fig:ALCSequentCalculusRules}
\end{figure}  

\begin{example}\textbf{(Sequent Proof for $\mathcal{ALC}$ Subsumption)}. Figure \ref{fig:ALCSequentProof} shows Example \ref{ex:QueryALC01}'s proof using the sequent calculus for $\mathcal{ALC}$. The cut rule is applied to the initial assumptions, according to theorem \ref{prop:CutEliminationTheorem}.
\end{example}

\begin{figure}[!htb]
\centering
{\small
\begin{prooftree}
\def\fCenter{\ \vdash\ }
  \Axiom$OL \fCenter \exists h.A \sqcap \forall h.C $
  \AxiomC{TRUE \quad} \RightLabel{=}
  \UnaryInf$A, C \fCenter C $ \RightLabel{l$\exists$} 
  \UnaryInf$\exists h.A , \forall h.C \fCenter \exists h.C$ \RightLabel{l$\sqcap$}  
  \UnaryInf$\exists h.A \sqcap \forall h.C \fCenter \exists h.C$ \RightLabel{cut}  
 \BinaryInf$OL \fCenter \exists h.C $ \RightLabel{cut}
  \Axiom$\exists h.C \fCenter CO$ \RightLabel{cut}
  \BinaryInf$(\exists h.C \fCenter CO , \; OL \fCenter \exists h.A \sqcap \forall h.C) \fCenter (OL \fCenter CO)$ \RightLabel{l$\sqcap$}
  
    \UnaryInf$\Big(\big((\exists h.C \fCenter CO) \sqcap (OL \fCenter \exists h.A \sqcap \forall h.C)\big) \fCenter (OL \fCenter CO)\Big)$  \RightLabel{l$\sqcap$}
\end{prooftree}
}
\caption{$\mathcal{ALC}$ sequent proof for $F_{1}$. The names of the clauses and the roles are abbreviated.}
\label{fig:ALCSequentProof}
\end{figure} 

This proof tree could be described by the following text in natural language: (1) If individuals who own at least one cat as a pet are owners of cats; and if the old ladies are, individuals who have at least one animal as a pet and all individuals who have only cat as pet. So this implies that old ladies own cats. (2) So, the old ladies are all people who have at least one cat as a pet. And all individuals who own at least one cat as a pet, own cats. (3) In addition to old ladies are all individuals who have at least one animal as a pet and all individuals who have only cat as pets; all individuals who have at least one animal as a pet and all individuals who have only cat as pets, are all individuals who have at least one cat 
as a pet. (4) Thus, an animal or a cat implies in a cat.

\section{Conversion Method}
\label{sec:MetodoConversao}

The process consists of two steps: building a formula tree and then converting this formula tree into sequents, given an $\mathcal{ALC}$ query and its matrix non-clausal connection proof. They are explained below.

\subsection{Building the Formula Tree}

\begin{figure}[!htb]
\centering
\subfloat[Internal Node]{
\includegraphics[height=1.7cm]{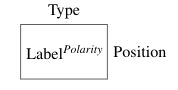}
\label{fig:NoInterno}.
}
\quad 
\subfloat[Leaf Node]{
\includegraphics[height=1.7cm]{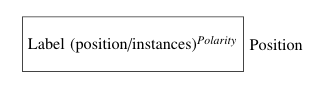}
\label{fig:nofolha}
}
\caption{Node Representation.}
\label{fig:TiposDeNos}
\end{figure}

\begin{definition}\label{def:arvore}\textbf{(Formula Tree, Position, Label, Polarity, Type).} A \textbf{formula tree} is a syntactic representation of a formula \textit{F} as a tree, where each node can have up to two child nodes. Each node has:


\textbf{Position:} an index that identifies each element (predicate or connective) in the formula. Its represented as $a_{0}, a_{1}, a_{2}, \ldots$; 
\textbf{Label:} either a connective $(\sqcap, \sqcup, \neg, \sqsubseteq, \models)$, quantifier or predicate, if it is an atomic (sub-)formula. Nodes whose label is a predicate are leaves of the tree (figure \ref{fig:nofolha}), while other nodes are  internal (figure \ref{fig:NoInterno}); 
\textbf{Polarity:} can be 0 or 1. It is determined by the label and the parent node polarity. The root node of the tree has polarity 0; 
\textbf{Type:} the type of a node is a Greek letter: $\alpha$, $\beta$, $\alpha'$, $\beta'$, $\gamma$ and $\delta$. It is determined by its label and its polarity. Leaf nodes have no type. The polarity and type of a node are defined in table \ref{tab:TCCCS-PolaridadeTipos}. For example, in the first line of this table, $(A \sqcap B)^{1}$ means that the node labelled $\sqcap$ and polarity 1 has type $\alpha$ and its successor nodes have polarity 1.
\end{definition}

\begin{table}[!ht]
\caption{Polarity and types of nodes for $\mathcal{ALC}$}
\label{tab:TCCCS-PolaridadeTipos}
\centering
\begin{tabular}{l||llll||llll||ll}

\cline{1-3}\cline{5-7}\cline{9-11}
Type $\alpha$ & \multicolumn{2}{l}{ }  &  & Type $\beta$ & \multicolumn{2}{l}{ }& &Type $\delta$ &\multicolumn{2}{l}{ }\\
\cline{1-3}\cline{5-7}\cline{9-11} 

$(A \sqcap B)^{1}$ & \multicolumn{1}{c}{$A^{1}$} & \multicolumn{1}{c}{$B^{1}$} &  &    
$(A \sqcap B)^{0}$ & \multicolumn{1}{c}{$A^{0}$} & \multicolumn{1}{c}{$B^{0}$} & &
$(\forall r A)^{0}$ & \multicolumn{1}{c}{$r^{1}$} & \multicolumn{1}{c}{$A^{0}$}  \\

$(A \sqcup B)^{0}$ & \multicolumn{1}{c}{$A^{0}$} & \multicolumn{1}{c}{$B^{0}$} &  & 
$(A \sqcup B)^{1}$ & \multicolumn{1}{c}{$A^{1}$} & \multicolumn{1}{c}{$B^{1}$} & &
$(\exists r A)^{1}$ & \multicolumn{1}{c}{$r^{1}$} & \multicolumn{1}{c}{$A^{1}$}  \\

$(\neg A)^{1}$  &  \multicolumn{2}{c}{$A^{0} \qquad$} &  &  &  &  & & & & \\   

$(\neg A)^{0}$  &  \multicolumn{2}{c}{$A^{1} \qquad$} &  &  &  &  & & & & \\   

\cline{1-3}\cline{5-7}\cline{9-11}
Type $\alpha'$ & \multicolumn{2}{l}{ }    &  & Type $\beta'$ & \multicolumn{2}{l}{ }& &Type $\gamma$ &\multicolumn{2}{l}{ }\\
\cline{1-3}\cline{5-7}\cline{9-11} 

$(A \sqsubseteq B)^{0}$ & \multicolumn{1}{c}{$A^{1}$} & \multicolumn{1}{c}{$B^{0}$} & & $(A \sqsubseteq B)^{1}$  & \multicolumn{1}{c}{$A^{0}$} & \multicolumn{1}{c}{$B^{1}$} & &
$(\forall r A)^{1}$ & \multicolumn{1}{c}{$r^{0}$} & \multicolumn{1}{c}{$A^{1}$}  \\

$(A \models B)^{0}$ & \multicolumn{1}{c}{$A^{1}$} & \multicolumn{1}{c}{$B^{0}$} & &  & \multicolumn{1}{c}{} & \multicolumn{1}{c}{} & &
$(\exists r A)^{0}$ & \multicolumn{1}{c}{$r^{0}$} & \multicolumn{1}{c}{$A^{0}$}  \\

\end{tabular}
\end{table}

Nodes of type $\alpha$ and $\alpha'$ correspond to sequent rules that do not cause proof branching. Nodes of type $\gamma$ and $\delta$ correspond to quantifier rules. Rules associated to type $\delta$ have the eigenvariable condition in the sequent calculi (where the term $t$, the eingevariable in the inference, appears in the main formula of inference and in no other formula in the sequent. In the case of the l$\exists$ rule for the existential quantifier and r$\forall$ rule for the universal quantifier). Nodes of type $\beta$ and $\beta'$ (i.e., $\sqcap^{0}$, $\sqcup^{1}$, and $\sqsubseteq^{1}$) are particularly important, since their respective rules in sequents
(described in table \ref{tab:TCCCS-PolaridadeTiposRegras}) split proof branching into two independent sub-proofs. Nodes have their types indexed in the formula tree to facilitate their identification, for example $\beta_{1}$, $\beta_{2}$, $\beta'_{1}$, $\beta'_{2}$. Each branch whose root is of type $\beta$ or $\beta'$ is marked with a letter (a,b,c,...).

Leaf nodes with instances are children of nodes type $\alpha$, $\alpha'$ or $\beta$. Leaf nodes without instances have labels attached to their closest predecessor nodes' position, according to the following criteria : (1) if the leaf node label represents a concept, it has an unique position associated to its label; (2) if the leaf node label represents a role, it has two positions associated to its label in the form ($a_{1},a_{2}$), where $a_{2}$ is the  of the nearest predecessor node's position; (3) only type $\gamma$, $\delta$ and $\beta'$ node positions are associated to the labels. This helps to check for complementarity in a connection between two nodes.

The tree construction is guided by the identification of the (sub-)formulae's main constructor (connective or quantifier), which will be a label in the tree node. This node has at most two branches that binds them to their child nodes, i.e., new (sub-)formulas. The node type and its children’s polarities are assigned according to table \ref{tab:TCCCS-PolaridadeTipos}. If children nodes are not atomic (sub-)formulae, the process repeats itself by identifying these (sub)-formulae's main constructor and then generating other nodes in the tree, until it reaches the leaves.

The proof matrix elements must correspond to the leaf nodes in the formula tree, indicated by the position of the corresponding predicate, as explained in section \ref{sec:TCCCS-ProcessoConversao} step 2.

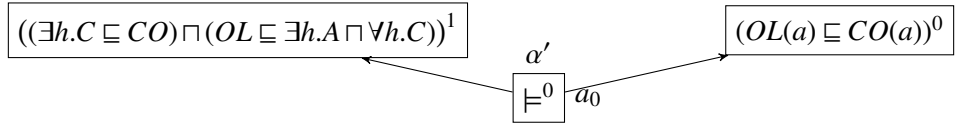
\begin{figure}[!htb]
\centering

\begin{tikzpicture}[grow=up,->,>=stealth']
\tikzstyle{level 1} = [sibling distance =8.0cm, level distance = 0.9cm]

\tikzstyle{aresta-r} = [font=\small, right]
\tikzstyle{aresta-l} = [font=\small, left]
   
\node [shape=rectangle, draw=black, label=right:$a_{0}$, label=above:$\alpha'$] {$\models^{0}$}
child{ node [shape=rectangle, draw=black, label=right:] {$(OL(a) \sqsubseteq CO(a))^{0}$}
    }
    child{ node [shape=rectangle, draw=black, label=left:] {$\big((\exists h.C \sqsubseteq CO) \sqcap (OL \sqsubseteq \exists h.A \sqcap \forall h.C)\big)^{1}$}
    }
; 
\end{tikzpicture}
\caption{Step 01 – Process of building the formula tree for $F_{1}$.}
\label{fig:Passo01-ArvoreFormula01}
\end{figure}

\begin{figure}[!htb]
\centering
\begin{tikzpicture}[grow=up,->,>=stealth']
\tikzstyle{level 1} = [sibling distance =5.5cm, level distance = 0.7cm]
\tikzstyle{level 2} = [sibling distance =3.9cm, level distance = 1.2cm]
\tikzstyle{level 3} = [sibling distance =2.0cm, level distance = 1.2cm]
\tikzstyle{level 4} = [sibling distance =4.0cm, level distance = 1.2cm]
\tikzstyle{level 5} = [sibling distance =2.3cm, level distance = 1.2cm]
\tikzstyle{level 6} = [sibling distance =1.5cm, level distance = 1.2cm]

\tikzstyle{aresta-r} = [font=\small, right]
\tikzstyle{aresta-l} = [font=\small, left]
   
\node [shape=rectangle, draw=black, label=right:$a_{0}$, label=above:$\alpha'$] {$\models^{0}$}
child{ node [shape=rectangle, draw=black, label=right:$a_{16}$, label=above:$\alpha'$] {$\sqsubseteq^{0}$}
   child{ node [shape=rectangle, draw=black, label=right:$a_{18}$] {$CO(a)^{0}$}} 
   child{ node [shape=rectangle, draw=black, label=right:$a_{17}$] {$OL(a)^{1}$}}                            
    }
    child{ node [shape=rectangle, draw=black, label=left:$a_{1}$, label=above:$\alpha$] {$\sqcap^{1}$}
	child{ node [shape=rectangle, draw=black, label=left:$a_{7}$, label=above:$\beta'_{2}$] {$\sqsubseteq^{1}$}
	child {node [shape=rectangle, draw=black, label=right:$a_{9}$, label=above:$\alpha$] {$\sqcap^{1}$}
	child{ node [shape=rectangle, draw=black, label=right:$a_{13}$, label=above:$\gamma$] {$\forall^{1}$}
	child{ node [shape=rectangle, draw=black, label=right:$a_{15}$] {$C(a_{13})^{1}$}}
	child{ node [shape=rectangle, draw=black, label=right:$a_{14}$] {$h(a_{7},a_{13})^{0}$}}}   		
	child {node [shape=rectangle, draw=black, label=right:$a_{10}$, label=above:$\delta$] {$\exists^{1}$}
	child {node [shape=rectangle, draw=black, label=left:$a_{12}$] {$A(a_{10})^{1}$}}
	child {node [shape=rectangle, draw=black, label=left:$a_{11}$] {$h(a_{7},a_{10})^{1}$}}}
	    edge from parent node[aresta-r] {d}
	}
	child {node [shape=rectangle, draw=black, label=right:$a_{8}$] {$OL(a_{7})^{0}$}
	      edge from parent node[aresta-l] {c}
	}}
	child{ node [shape=rectangle, draw=black, label=left:$a_{2}$, label=above:$\beta'_{1}$] {$\sqsubseteq^{1}$}
	child{ node [shape=rectangle, draw=black, label=left:$a_{6}$] {$CO(a_{2})^{1}$}
	   edge from parent node[aresta-r] {b}
	}
	child{ node [shape=rectangle, draw=black, label=left:$a_{3}$, label=above:$\gamma$] {$\exists^{0}$}
	child {node [shape=rectangle, draw=black, label=left:$a_{5}$] {$C(a_{3})^{0}$}
	}
	child {node [shape=rectangle, draw=black, label=left:$a_{4}$] {$h(a_{2},a_{3})^{0}$}
	}
	    edge from parent node[aresta-l] {a}	 
	}}
    }
; 
\end{tikzpicture}
\caption{Formula Tree for $F_{1}$ with labels, polarities and types.}
\label{fig:ArvoreFormula01}
\end{figure}
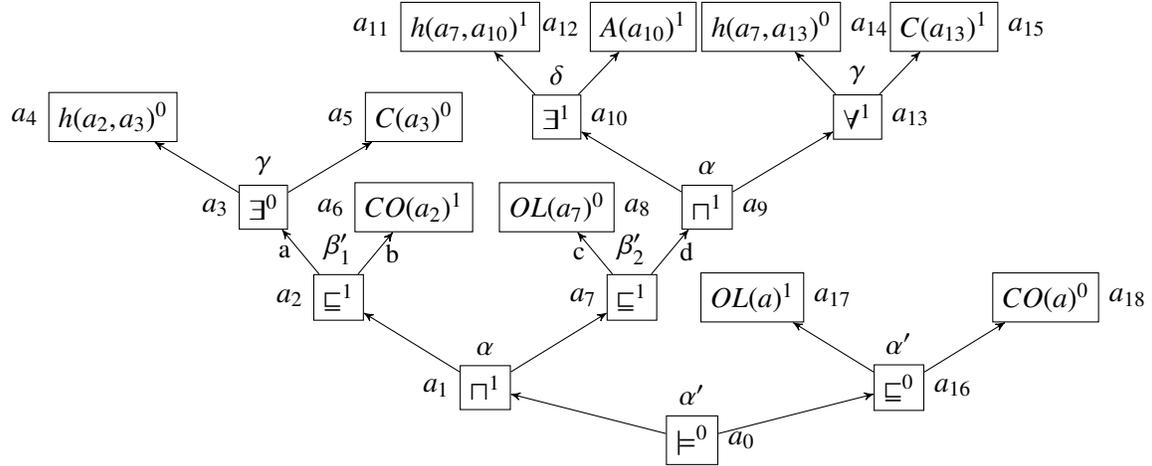

\begin{example}\textbf{(Building the Formula Tree Process).} Figure \ref{fig:Passo01-ArvoreFormula01} shows the first step in the tree construction for $F_{1}$ from Example \ref{ex:QueryALC01}: $\big((\exists h.C \sqsubseteq CO) \sqcap (OL \sqsubseteq \exists h.A \sqcap \forall h.C)\big) \models \big(OL(a) \sqsubseteq CO(a)\big)$. Its main constructor is $\models$, the root node label, which, by definition has polarity 0; its position is $a_{0}$. According to table \ref{tab:TCCCS-PolaridadeTipos}, its type is $\alpha'$; its children nodes, on the right and left, have polarities 0 and 1, respectively, and both are sub-formulas of $\models$ in $F_{1}$. This process continues until it reaches the leaf nodes, as shown in figure \ref{fig:ArvoreFormula01}.
\end{example}

For a given formula $A$, $A'$, $B$, $B'$, $\Gamma$ and $\Delta$ are used to denote the sets of node positions of type $\alpha$, $\alpha'$, $\beta$,  $\beta'$, $\gamma$, and $\delta$, respectively.

\begin{definition}\textbf{(Substitution of positions $\sigma_{\delta}$, ordering relation $\sqsubset_{\delta}$))}. It replaces positions of type $\gamma$ for positions of type $\delta$. A \textbf{position substitution $\sigma_{\delta}$} is a mapping of the set $\Gamma$ of type $\gamma$ node positions to the set $\Delta$ of type $\delta$ node positions. The $\sigma_{\delta}$ substitution induces a \textbf{partial ordering relation $\sqsubset_{\delta}$} in $\Delta \times \Gamma$ as follows: let $u \in \Gamma$ and $v \in \Delta$; if $\sigma_{\delta}(u) = p$ then $v \sqsubset_{\delta} u$ for all $v \in \Delta$ occurring in position $p$.
\end{definition}

Since the sequent rules $r \forall$ and $l \exists$ and their homologues $l \neg \forall$ and $r \neg \exists$ are restricted to the eigenvariable condition, the relation $v \sqsubset_{\delta} u$ expresses that the node labelled by $v$ must be reduced before reducing the one labelled by $u$.

\begin{example}\textbf{(Substitution of positions $\sigma_{\delta}$, ordering relation $\sqsubset_{\delta}$)}. Consider the formula tree in figure \ref{fig:ArvoreFormula01}. Let $u$ be the node labelled by $\forall^{1}$, with position $a_{13}$ and type $\gamma$, and let $v$ be the node labelled by $\exists^{1}$, with position $a_{10}$ and type $\delta$. To replace the position of a Type $\gamma$ node by the position of a type $\delta$ node, It is necessary to reduce the type $\delta$ node first, then the node with the position $a_{10}$ must be reduced before the node with the position $a_{13}$. Thus, for this example, the ordering relation $\sqsubset_{\delta}$ is given by $\exists^{1}a_{10} \sqsubset_{\delta} \forall^{1}a_{13}$, and the substitution $\sigma_{\delta}(\forall^{1}a_{13}) = a_{10}$. With this, we have $\sigma_{\delta}=\lbrace a_{13} / a_{10} \rbrace$.
\end{example}

\begin{definition}\textbf{(Substitution of positions $\sigma_{\beta'}$).} It replaces positions of type $\beta'$, $\gamma$, $\delta$ for instances or positions of type $\beta'$. Positions of the nodes of type $\beta'$, $\gamma$ and $\delta$, as well as instances, appear in atomic formulas, so a \textbf{substitution of positions $\sigma_{\beta'}$} is a mapping of the set $B'/\Gamma/\Delta$ positions of nodes of type $\beta'/\gamma/\delta$ to instances or positions of nodes of type $\beta'$. Let $u$ be a leaf node with the positions of nodes of type $\beta'/\gamma/\delta$ associated to its label and $v \in B'$; if $\sigma_{\beta'}(u) = p$, where $p \in B'$ or $p$ is an instance.
\end{definition}

Reducing a node means applying the sequent rule that corresponds to that node over a given (sub\nobreakdash-)for-mula. Leaf nodes are not reduced.

\begin{example} \textbf{(Substitution of positions $\sigma_{\beta'}$)}.  Consider the formula tree in Figure \ref{fig:ArvoreFormula01}. Let $u$ be the node labelled by $OL^{0}$, with position $a_{8}$ and position $a_{7}$ of type $\beta'$ associated to its label, and let $v$ be the node labelled by $OL^{1}$, with position $a_{17}$ and instance $a$. The substitution for this in leaf $u$ in this case is $\sigma_{\beta'}(OL(a_{7})^{0}) = a$. Therefore, $\sigma_{\beta'}=\lbrace a_{7} / a \rbrace$.
\end{example}

\begin{definition}\textbf{(Substitution $\sigma_{Final}$)}. It is a combination of $\sigma_{\delta}$ and $\sigma_{\beta'}$. A \textbf{$\sigma_{Final}$ substitution} consists of a substitution $\sigma_{\delta}$ and a substitution $\sigma_{\beta'}$, where $\sigma_{Final} := \sigma_{\delta} \cup \sigma_{\beta'}$.
\end{definition}

\begin{example}\textbf{(Substitution $\sigma_{Final}$)}. Considering the two previous examples, $\sigma_{Final} = \lbrace a_{13} / a_{10} , a_{7} / a \rbrace$. 
\end{example}

\begin{definition}\textbf{(Connection, $\sigma_{Final}$-complementary connection)}. A \textbf{connection} is a pair of leaf nodes labelled with the same predicate symbol and the same position associated with the label or the same instance, but with different polarities. If they are identical under $\sigma_{Final}$, the connection is a \textbf{$\sigma_{Final}$-complementary connection}.
\end{definition}

\begin{example} \textbf{(Connection, $\sigma_{Final}$-complementary connection)}. Let the formula tree in figure \ref{fig:ArvoreFormula01} be. The leaf nodes $h(a_{2},a_{3})^{0}$ and $h(a_{7},a_{10})^{1}$ with positions $a_{4}$ and $a_{11}$, respectively, form a connection that is complementary under $\sigma_{Final} = \lbrace a_{2} / a_{7} , a_{3} / a_{10} \rbrace$.
\end{example}

\begin{definition}\textbf{(Tree Ordering $\prec$).} The \textbf{tree ordering $\prec$} of an $F$ formula is the partial ordering of the nodes positions in the tree formula. $\prec$ is defined as follows:\textit{(i)} the root occupies the smallest position with respect to this ordering, \textit{(ii)} $a_{i} \prec a_{j}$ if and only if the position $a_{i}$ is below $a_{j}$ in the formula tree.
\end{definition}

\begin{example} \textbf{(Tree Ordering $\prec$)}. In the tree from Figure \ref{fig:ArvoreFormula01}, there are examples of tree ordering: $a_{7} \prec a_{9} \prec a_{13} \prec a_{15}$ and $a_{0} \prec a_{1} \prec a_{2} \prec a_{3}$.
\end{example}

\begin{definition}\textbf{(Reduction Order $\lhd$).} The transitive closure of the union of $\sqsubset_{\delta}$, $\sqsubset_{\beta'}$ and $\prec$ is called \textbf{reduction order $\lhd$}, i.e., $\lhd := (\prec \cup \sqsubset_{\delta} \cup \sqsubset_{\beta'})^{+}$.
\end{definition}

Nodes $v \lhd u$ means that the node $v$ must be reduced before the node labelled by $u$ in the sequent poof. $\lhd$ determines the nodes' reduction order, and helps determine which sequent rules are to be used and in which order.

\begin{example} \textbf{(Reduction Order $\lhd$)}. In Figure \ref{fig:ArvoreFormula01}, the nodes with positions $a_{7}$, $a_{10}$, $a_{16}$ and $a_{13}$, have the following reduction order $\lhd$: \textit{(i)} $a_{7} \prec a_{10}$; \textit{(ii)} $a_{7} \prec a_{13}$; \textit{(iii)} $a_{10} \sqsubset_{\delta} a_{13}$. The orderings' union and the tree ordering determine the reduction order for these nodes: $ a_{7} \lhd a_{10} \lhd a_{13}$.
\end{example}

\begin{definition}\textbf{($\sigma_{Final}$ Admissible Substitution).} An \textbf{$\sigma_{Final}$ Substitution is admissible} if the reduction order $\lhd$ is not  reflexive. In this case, it is possible to construct a sequent proof.
\end{definition}

A correspondence between node label, polarity and type with the sequent rules presented in section \ref{sec:SequentCalculusALCSubsumption}, is established in table \ref{tab:TCCCS-PolaridadeTiposRegras}. Such correspondence is useful for the sequent proof construction, where the polarity helps in the identification of the rule. Polarity 1 represents a rule on the left (left or l); polarity 0, on the right (right or r), for cases where there is already an associated rule. For instance, in Table \ref{tab:TCCCS-PolaridadeTiposRegras}'s first line, for node $\sqcap^{1}$ the rule is l$\sqcap$, while for node $\sqcap^{0}$ it is r$\sqcap$. For cases where internal nodes are preceded by a node labelled by a negation, correspondences are in Table \ref{tab:TCCCS-PolaridadeTiposRegras}'s last four columns.

\begin{table}[ht]
\caption{Correspondence between label, polarity and type of a node, preceded or not by a node labelled with negation, to $\mathcal{ALC}$ Sequent rules.}
\label{tab:TCCCS-PolaridadeTiposRegras}
\centering
{\small
\begin{tabular}{|l|l|l|l|l|l|l|l|l|l|l|l|l|l|}
\cline{1-8}\cline{10-14}
\multicolumn{8}{|c|}{Not preceded}   & & \multicolumn{5}{|c|}{Preceded}       \\ 
\cline{1-2}\cline{4-5}\cline{7-8}\cline{10-11}\cline{13-14}
Type $\alpha$ & Rule  &  & Type $\beta$ & Rule & &Type $\delta$ & Rule & & Type $\alpha$ & Rule & & Type $\beta$ & Rule \\
\cline{1-2}\cline{4-5}\cline{7-8}\cline{10-11}\cline{13-14}

$\sqcap^{1}$ & l$\sqcap$ &  & $\sqcap^{0}$ & r$\sqcap$ & &
$\forall^{0}$ & r$\forall$ & & $\neg^{1}$ & r$\neg\neg$ & &$\sqcap^{0}$ & l$\neg\sqcap$ \\

$\sqcup^{0}$ & r$\sqcup$ &  & $\sqcup^{1}$ & l$\sqcup$ &  &
$\exists^{1}$ & l$\exists$  & & $\neg^{0}$& l$\neg\neg$& & $\sqcup^{1}$ & r$\neg\sqcup$ \\

$\neg^{1}$  & $\varnothing$ &  &  &  &  &  & &  & $\sqcap^{1}$ & r$\neg\sqcap$ & & &\\  

$\neg^{0}$  & $\varnothing$ &  &  &  &  &  &  & & $\sqcup^{0}$ & l$\neg\sqcup$ & & &\\   

\cline{1-2}\cline{4-5}\cline{7-8}\cline{10-11}\cline{13-14}
Type $\alpha'$ & Rule &  & Type $\beta'$ & Rule & & Type $\gamma$ & Rule &  & & & & Type $\delta$ & Rule\\
\cline{1-2}\cline{4-5}\cline{7-8}\cline{10-11}\cline{13-14}

$\sqsubseteq^{0}$ & $\varnothing$ &  & $ \sqsubseteq^{1}$  & Cut &  &
$\forall^{1}$ & $\varnothing$ & & & & & $\forall^{0}$ &l$\neg\forall$\\

$\models^{0}$ & $\varnothing$ &  &  &  &  & $\exists^{0}$ & $\varnothing$  & & & & & $\exists^{1}$ &r$\neg\exists$ \\
\cline{1-8}\cline{10-14}
\end{tabular}
}
\end{table}

\subsection{Conversion to Sequents}
\label{sec:TCCCS-ProcessoConversao}

Given an $\mathcal{ALC}$ query and its matricial non-clausal connection proof, the conversion procedure transforms this proof into an $\mathcal{ALC}$ sequent proof. This process performs four steps, which are described below:

\begin{itemize}
\item \textbf{Step 1- Formula tree construction:} A syntactic representation in tree form is constructed for the input formula, containing nodes, as described in \ref{def:arvore}. The position of each predicate is input to step 2, and the tree to steps 3 and 4. \textbf{Example:} The conversion process begins with the $F_{1}$ formula tree construction, described in definition \ref{def:arvore}, which resulted in the formula tree represented in figure \ref{fig:ArvoreFormula01}.
\item \textbf{Step 2- Matrix elements' positions assignment:} Since proof matrix elements correspond to predicates in the formula and also to leaf nodes in the formula tree, this step assigns to each matrix element the position of the corresponding predicate. Its input is the matrix non-clausal connection proof and the position of predicates. Its output is input to step 3. \textbf{Example:} Each element of the matrix is assigned with the position of the corresponding predicate in the formula, see matrix in \ref{fig:PassosConversao}.
 
\begin{figure}[!htb]
\includegraphics[height=5.7cm, width=13cm]{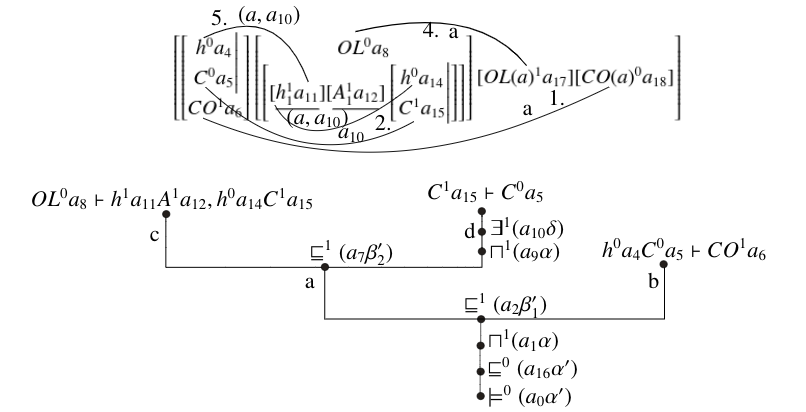}
\caption{Steps representation in the connection proof/sequent $\mathcal{ALC}$ for $F_{1}$.}
\label{fig:PassosConversao}
\end{figure}
 
\item \textbf{Step 3- (partial) sequent proof structure Construction:} The matrix non-clausal connection proof with the positions of each element and the formula tree are inputs for this step. To each matrix connection, the formula tree is examined in search for the leaf nodes that correspond to the connection. The paths between the root node and these nodes in the tree are analyzed to determine the order of nodes to be worked on and thus build a structure of the (partial) proof in sequents. This structure provides information about the reduction order $\lhd$, which helps determine the rules to be applied, and on the existence of the proof branch, given by the identification of the nodes of type $\beta$ and $\beta'$. The (partial) sequent proof structure constructed will be the input for step 4. \textbf{Example:} The first connection links element $CO(a)^{0}$, from position $a_{18}$, to element $CO^{1}$, of position $a_{6}$, which are complementary under the substitution $\sigma_{\beta'} = \lbrace a_{2} / a \rbrace$, see table \ref{tab:conexaoSubstOrdem04}. The path between these leaf nodes is $\lbrace a_{18}, a_{16}, a_{0}, a_{1}, a_{2}, a_{6} \rbrace$. Since there is no ordering relation $\sqsubset_{\sigma}$ between the nodes of that path and there are two tree orderings given by $a_{0} \prec a_{16}$ and $a_{0} \prec a_{1} \prec a_{2}$, It is possible to start with any of these tree orderings. Choosing the first, we have the order of reduction at that moment equal to: $a_{0} \lhd a_{16} \lhd a_{1} \lhd a_{2}$. Since the node with position $a_{2}$ is of type $\beta'$, the sequent is divided into two branches, called ’a’ e ’b’, as in the formula tree. Thus, this connection closes the branch ’b’, branch where the node $CO^{1}$ is, and leads to the axiom $h^{0}, C^{0} \vdash CO^{1}$, because nodes of type $\beta'$ are associated with the cut rule (see table \ref{tab:TCCCS-PolaridadeTiposRegras}). In the second connection, $\underline{C^{0}}$, with position $a_{5}$ in branch ’a’, is connected to $\underline{C^{1}}$, with position $a_{15}$ in branch ’d’, and the path between them is $\lbrace a_{5}, a_{3}, a_{2}, a_{1}, a_{7}, a_{9}, a_{13}, a_{15}\rbrace$. As the nodes with positions $a_{1}$ and $a_{2}$ have already been reduced, it is necessary to reduce the nodes with positions, $a_{3}$, $a_{7}$, $a_{9}$ and $a_{13}$, which have tree ordering $a_{7} \prec a_{9} \prec a_{13}$ and the relations $a_{10} \sqsubset_{\delta} a_{3}$ and $a_{10} \sqsubset_{\delta} a_{13}$. At the moment it is only possible to reduce the node with position $a_{7}$ and then the node with position $a_{9}$, that is, $a_{7} \lhd a_{9}$. Since the node with position $a_{7}$ is of type $\beta'$, its reduction divides branch 'a' into branches 'c' and 'd'. Then the node with position $a_{9}$, in branch 'd', is reduced. Since there are pendant nodes on this path, it is not yet possible to form an axiom and close the 'd' branch. The third connection is analyzed, where $\underline{h^{0}}$, with position $a_{14}$, is connected to $\underline{h^{1}}$, with position $a_{11}$, both in branch 'd'. The path between the nodes is $\lbrace a_{14}, a_{13}, a_{9}, a_{10}, a_{11}\rbrace$. Since $a_{9}$ has already been reduced, and there are the relations $a_{10} \sqsubset_{\delta} a_{13}$ and $a_{10} \sqsubset_{\delta} a_{3}$, the $a_{10}$ position node is reduced, and 'together' with it the nodes with position $a_{13}$ and $a_{3}$. The reduction of the $a_{10}$ position node makes the third and second connection complementary under the substitutions $\sigma_{\delta} = \lbrace a_{13} / a_{10}, \; a_{3} / a_{10} \rbrace$. With this the last two connections are reflected in the sequent proof leading to the closure of the 'd' branch. Notice that the second connection was only reached in the tree after the third connection, this leads to the axiom in the form $C^{1} \vdash C^{0}$. The fourth connection connects $OL^{0}$, with position $a_{8}$ in branch 'c', to $OL^{1}$, with position $a_{17}$. The path between the nodes with theses positions is $\lbrace a_{8}, a_{7}, a_{1}, a_{0}, a_{16}, a_{17}\rbrace$. As all nodes on this path have already been reduced, no reduction will be necessary in this step. Thus, 'c' branch is closed with an axiom in the form $OL^{0} \vdash h^{1}A^{1},h^{0}C^{1}$, due to the cut rule. This connection is complementary under $\sigma_{\beta'} = \lbrace a_{7} / a\rbrace$. On the fifth and last connection, which connects $\underline{h^{0}}$ to $\underline{h^{1}}$, there is no need of node reduction, since all nodes in the path were reduced. The connection is complementary under $\sigma_{\delta} = \lbrace a_{3} / a_{10}\rbrace$. Note that $a_{2} / a$ and $a_{7} / a$ were $\sigma_{\beta'}$ previous substitutions. All connections are complementary under a substitution $\sigma_{Final}$, all branches of the proof structure in sequent were closed, and the reduction order is not reflexive, as shown in figure \ref{fig:PassosConversao} and in Table \ref{tab:conexaoSubstOrdem04}.

\begin{table}[ht]
\centering
\caption{Relation between connections, substitutions and orderings}
\label{tab:conexaoSubstOrdem04}
\begin{tabular}{|l|l|l|l|l|l|}
\hline
Nº & Nodes & $\sigma_{\delta}$ & $\sigma_{\beta'}$ & $\sqsubset_{\delta}$ & $\lhd$ \\ \hline
1  & $CO(a_{2})^{1}a_{6},CO(a)^{0}a_{18}$ & & $a_{2}/a$ &   & $a_{0} \lhd a_{16} \lhd a_{1} \lhd a_{2}$ \\ \hline
2  & $C(a_{3})^{0}a_{5},C(a_{13})^{1}a_{15}$  & \begin{tabular}[c]{@{}l@{}}$a_{13}/a_{10}$,\\$a_{3}/a_{10}$\end{tabular} & &
\begin{tabular}[c]{@{}l@{}}$a_{10} \sqsubset_{\delta} a_{3}$,\\ $a_{10} \sqsubset_{\delta} a_{13}$ \end{tabular} & $a_{7} \lhd a_{9}$                        \\ \hline
3  & $h(a_{7},a_{13})^{0}a_{14},h(a_{7},a_{10})^{1}a_{11}$ & $a_{13}/a_{10}$  &   &   & $a_{10}$  \\ \hline
4  & $OL(a_{7})^{0}a_{8},OL(a)^{1}a_{17}$ &   & $a_{7}/a$ &     &  \\ \hline
5  & $h(a_{2},a_{3})^{0}a_{4},h(a_{7},a_{10})^{1}a_{11}$      & $a_{3}/a_{10}$               & $a_{2}/a$               &                  &                                           \\ \hline
\multicolumn{3}{|l|}{$\sigma_{Final}= a_{2}/a, \; a_{13}/a_{10}, \; a_{3}/a_{10}, \; a_{7}/a$} &
\multicolumn{3}{|l|}{$a_{0} \lhd a_{16} \lhd a_{1} \lhd a_{2} \lhd a_{7} \lhd a_{9} \lhd a_{10}$} \\ \hline
\end{tabular}
\end{table}

\item \textbf{Step 4- Construction of the complete sequent proof}: Here, the process builds a complete sequent proof (output) from the (partial) sequent proof structure and the correspondence between nodes and sequent rules, described in \ref{tab:TCCCS-PolaridadeTiposRegras}. The input is (partial) sequent proof structure, the formula tree and $\mathcal{ALC}$ sequent rules. \textbf{Example:} The structure obtained in step 3 is traversed. The proof begins with the reduction of $a_{1}$ position node, since the first two tree nodes do not have associated rule, because they are of type $\alpha'$. Rule l$\sqcap$ is applied. Then, the $a_{2}$ position node, with type $\beta'$, é reduced by means of the cut rule on the query $\alpha$, that is, on $(OL \vdash CO)$. The proof is divided into branches 'a' and 'b'. The 'b' branch is closed with the initial axiom $\exists h.C \vdash CO$, while branch 'a' is open, in which $OL \vdash \exists h.C$ must be proved. The next node is of position $a_{7}$, of type $\beta'$, and its reduction divides the branch 'a' into branches 'c' and 'd', by means of the application of a new cut rule on $OL \vdash \exists h.C$. The 'c' branch is closed with the initial axiom $OL \vdash \exists h.A \sqcap \forall h.C $, while the 'd' branch stays open. To close the 'd' branch, the $a_{9}$ position node is reduced with the $l \sqcap$ rule, followed by the node with position $a_{10}$, through rule $l\exists$. This ends the $F_{1}$ sequent proof, as shown in figure \ref{fig:CCS-ProvaSequenteCompleta}:  
\end{itemize}

\begin{figure}[!htb]
\centering
{\small
\begin{prooftree}
\def\fCenter{\ \vdash\ }
  \Axiom$OL \fCenter \exists h.A \sqcap \forall h.C $
  \AxiomC{} \RightLabel{=}
  \UnaryInf$A, C \fCenter C $ \RightLabel{l$\exists$} 
  \UnaryInf$\exists h.A , \forall h.C \fCenter \exists h.C$ \RightLabel{l$\sqcap$}  
  \UnaryInf$\exists h.A \sqcap \forall h.C \fCenter \exists h.C$ \RightLabel{cut}  
 \BinaryInf$OL \fCenter \exists h.C $ \RightLabel{cut}
  \Axiom$\exists h.C \fCenter CO$ \RightLabel{cut}
  \BinaryInf$(\exists h.C \fCenter CO , \; OL \fCenter \exists h.A \sqcap \forall h.C) \fCenter (OL \fCenter CO)$ \RightLabel{l$\sqcap$}
  
    \UnaryInf$\Big(\big((\exists h.C \fCenter CO) \sqcap (OL \fCenter \exists h.A \sqcap \forall h.C)\big) \fCenter (OL \fCenter CO)\Big)$  \RightLabel{l$\sqcap$}
\end{prooftree}
}
\caption{Complete proof in $\mathcal{ALC}$ sequents for $F_{1}$.}
\label{fig:CCS-ProvaSequenteCompleta}
\end{figure} 
\section{Complexity} 
\label{chp:AlgoritmoComplexidade}

This section presents a very brief overview of the main algorithms for the conversion method with its complexities, according to the 4 steps seen in section \ref{sec:TCCCS-ProcessoConversao}. All the algorithms are demonstrated
in \cite{Palmeira:2}. Time complexities were analyzed according to the input size of each algorithm. For example, some algorithms receive an $\mathcal{ALC}$ formula $F$ as input, so the input size $n$ represents the number of symbols of $F$. Other algorithms accept an $F$ proof matrix as input; in this case, the input size is the matrix number of symbols, including connections between literals. This input is represented by $m$.

\begin{figure}[!htb]
\centering
\includegraphics[scale=0.5] {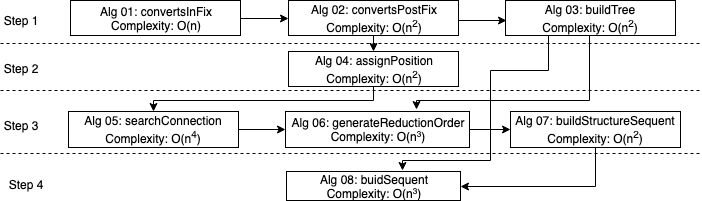}
\caption{Overview of the main algorithms' order.}
\label{fig:algoritmos}
\end{figure}

Figure \ref{fig:algoritmos} presents the main algorithms' execution order. Lines with arrows indicate that the output of one algorithm is input to another. For example, the output from algorithm 02 (called \textit{convertsPostFix}) is conveyed as input for algorithm 03 (called \textit{buildTree} and 04 (called \textit{assignPosition}). The complexity of algorithm 05 (\textit{Search Connections}) is the highest among the algorithms: O($n^{4}$), up to four iterations over structures based on the input size $m$.


\section{Conclusions}
This work presents a method to convert Non-clausal $\mathcal{ALC}$ connections proofs into more readable proofs. The approach consists in transforming these proofs into proofs in the $\mathcal{ALC}$-Sequent Calculus \cite{Borgida:1}. Hence, this conversion assumes that the input formulae will always be in non-clausal form, i.e., without the need to transform these formulae into any normal form. 
A tree representation of formulae is used as a guide in this conversion and a sequent proof is created while the connection proof is traversed. This conversion must contribute to describe how the reasoners based on the $\mathcal{ALC}$ Connection Method summon their inferences and may facilitate the creation of natural language explanations, given the ease of converting sequents to texts. The evaluation of the main algorithms' computational complexities demonstrates its practical feasibility, since they display polynomial complexity. In this perspective, the scientific contributions of this work should characterize the importance of the logical proofs, clarify the reasoning process and increase inferences' readability, thus providing better user interaction with connection reasoners.



\bibliographystyle{eptcs}
\bibliography{generic}

\end{document}